\documentclass[12pt,a4paper]{article}
\usepackage{graphicx}
\usepackage{times}
\title{\bf A multiwavelength approach to  BD+53$^\circ$2790:  the  O9.5V  counterpart  to the  X-ray  binary  system   4U~2206+54}
\author{Pere Blay$^1$ and V\'\i ctor Reglero$^1$\\
\vspace{1cm}\\
\normalsize $^1$ IPL, University of Valencia, PO BOX 22085, 46071 Valencia, Spain\\} 
\date{\mbox{}}
\begin{document}
\maketitle
\pagestyle{empty}
%
%
%
%
\def\beginrefer{\section*{References}%
\begin{quotation}\mbox{}\par}
\def\refer#1\par{{\setlength{\parindent}{-\leftmargin}\indent#1\par}}
\def\endrefer{\end{quotation}}
%
%
{\noindent\small{\bf Abstract:} 
The X-ray binary system 4U 2206+54 hides many mysteries. Among them,  the surprising behavior of both of its 
components: the O9.5 dwarf star BD+53$^\circ$2790  and a slowly rotating neutron star. BD+53$^\circ$2790 
misled the astronomers  showing itself very likely as a Be star. However, a deeper spectral analysis  and 
more intense monitoring, revealed that the real picture was a bit more  complicated: a) Although it shows 
evidence of a circumstellar envelope, its  observable properties differ from those typical envelopes in 
Be stars. b)  Comparison with spectral standards and models indicates a possible over-abundance  in He. This 
would open the possibility to link the behavior of  BD+53$^\circ$2790 to  the He-rich class of stars. c) UV 
spectra shows an abnormally slow and dense  wind for an O9.5V star. d) Spectral classification in the IR 
wavelength   region suggest a more likely supergiant  nature of the source, in contradiction with the  
optical classification. e) The presence of an intense magnetic field is under  investigation. 
BD+53$^\circ$2790 stands as a perfect laboratory for testing stellar structure, as well as  wind and 
evolutionary theories. The observable properties of this source in a wide range of spectral bands are 
discussed, and some interpretations outlined.
}
%
%
\section{Introduction}
X-ray binary systems consist of two stellar bodies orbiting each other and emitting high energy photons 
as the result of mass transfer from one onto the other. BD+53$^\circ$2790 is the massive companion of 
one of such systems, namely 4U 2206+54.  Although initially BD+53$^\circ$2790 was mis-classified as a 
Be star, Negueruela \& Reig (2001) and Blay et al. (2006) show reasons to discard this classification 
and suggest that BD+53$^\circ$2790 is instead a very peculiar main sequence O9.5 star. We will briefly 
review its peculiarities in this poster, from a wide point of view only achieved with a  multiwavelength 
approach. In X-ray binary systems, each part of the electromagnetic spectrum gives us information from  
a very specific part of the system. While optical data shows us what kind of massive companion is 
contained in 4U 2206+54, as well as its stellar parameters and properties, Infra-Red data indicates the 
properties of the very close neighborhood of this massive star, UV bands tell us how the mass-transfer 
takes place, and high energy and radio data helps to unveil what kind of accretor is hosting 4U 2206+54. 
Only a complete multiwavelength approach will help to comprehend BD+53$^\circ$2790 and its compact companion.

\section{High energy data: The compact companion}
\begin{figure}[h]
\centering
\begin{minipage}{12cm}
 \centering
\includegraphics[width=12cm]{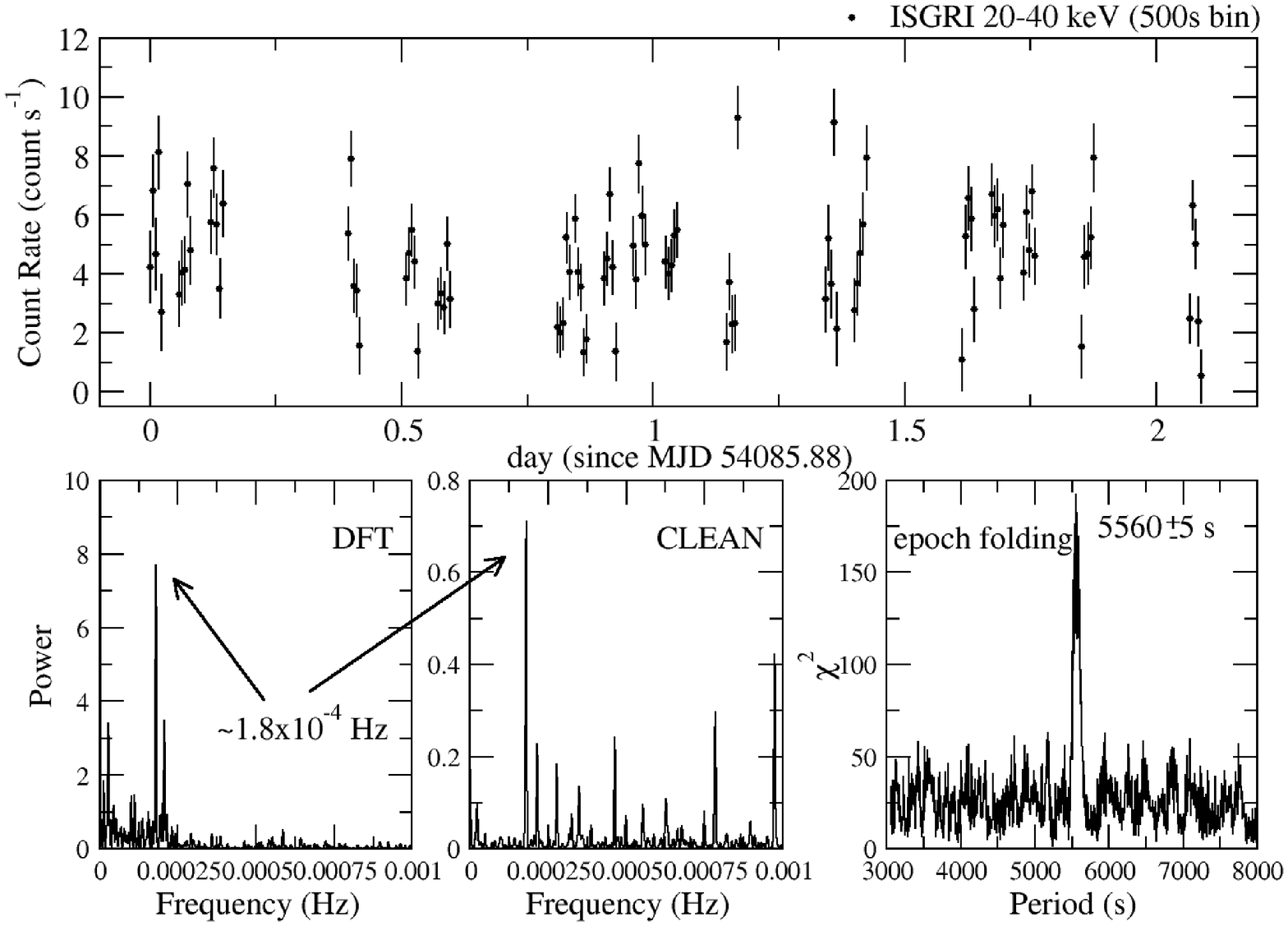}
\end{minipage}
\begin{minipage}{14cm}
 \centering
\includegraphics[width=14cm]{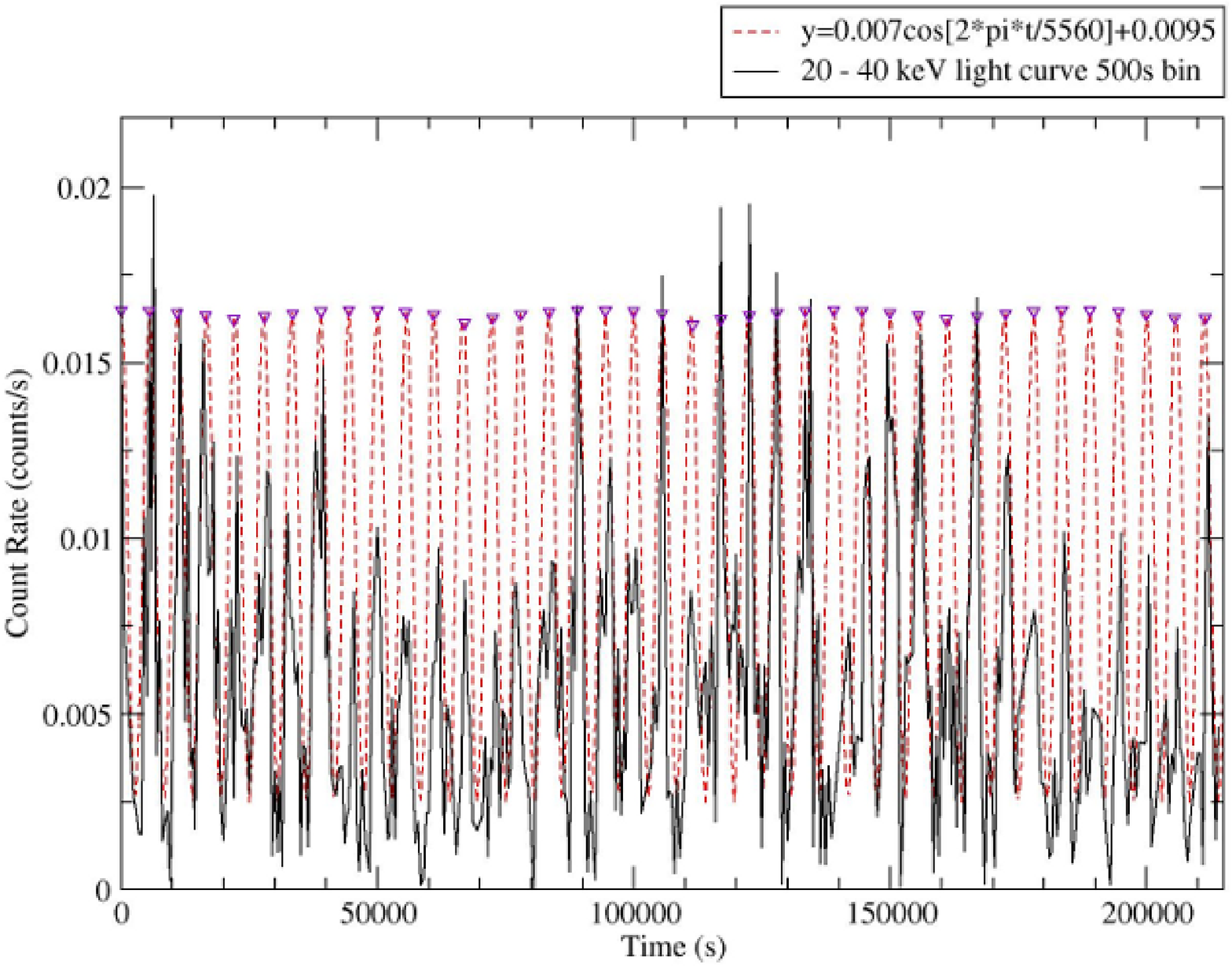}
\end{minipage}
\caption{\textbf{Upper panels}:  500s binned light curve (upper plot) and 
periodograms  (lower  plots) showing the pulse period 
detection from  4U 2206+54 by  INTEGRAL/ISGRI 
data in the 20-40 keV  energy range.  A clear 
periodic variation at 1.8·10-4 Hz (5560s) is shown. \textbf{Lower panel}: Pulsed emission from 4U 2206+54 compared to a
sinusoidal-like variation with a 5560s periodicity.  The
irregularity of the pulse profile is evident and striking.\label{fig_1}}
\end{figure}

\begin{figure}[h]
\centering
\includegraphics[width=12cm]{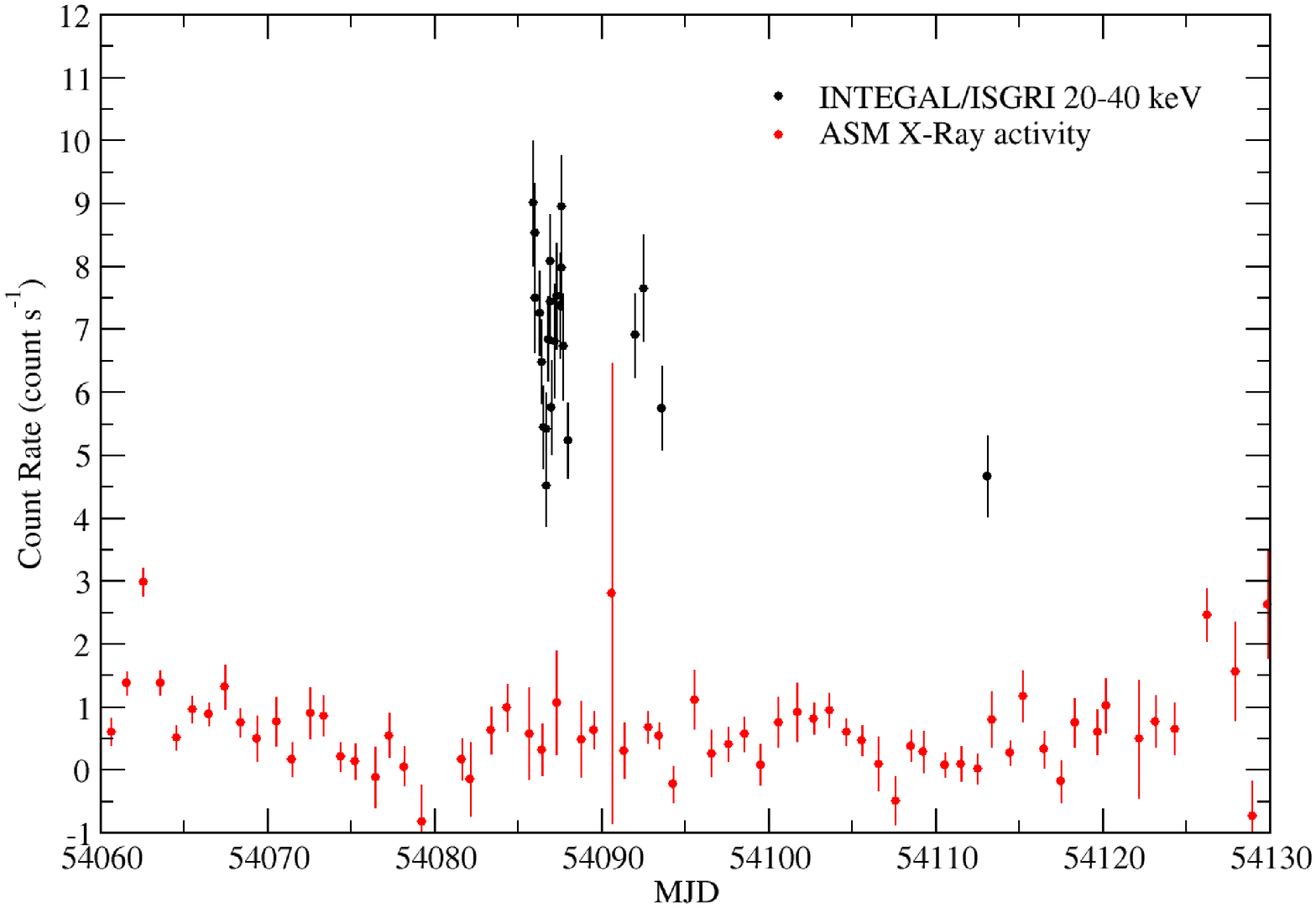}
\caption{X-ray light curve of 4U 2206+54 in the 2-10 keV 
 (RXTE/ASM) and 20-40 keV (INTEGRAL/ISGRI) energy
ranges. The flaring-like variability typical of wind fed
HMXRBs is shown in both energy ranges.\label{fig_2}}
\end{figure}

X and $\gamma$-ray measurements lead to the conclusion that 4U 2206+54 hosts a neutron star. Its pulse 
period has only been detected recently, by Reig et al. (2009) and 
Finger et al. (2010), to be 5560s. It is one of the longest pulse periods for a 
High Mass X-ray Binary System (HMXRB).  INTEGRAL detection of the pulse period is shown on the two 
plots of Figure \ref{fig_1}. The high energy light-curve of 4U 2206+54 (Figure \ref{fig_2}) is 
reminiscent of that of a HMXRB hosting a supergiant star and a wind-fed accreting compact companion 
(see Rib\'o et al. 2006).

Although a 9.56d periodicity was previously considered as the orbital period of the system 
(see Rib\'o et al. 2006), more recent high energy data indicates that most likely the 
orbital period is almost double this value, namely 19.25d (see Corbet, Markwardt, \& Tueller 2007).

\section{UV data: }
There are very few data in the UV range for BD 53$^\circ$2790.  Two low resolution and one high resolution 
spectra are available in the IUE (International Ultraviolet Explorer) database. They are described in 
detail in Negueruela \& Reig. (2001) and Blay, Rib\'o, \& Negueruela (2009). Rib\'o et al. (2006) found that the 
stellar wind from BD+53$^\circ$2790  was slower and denser than expected for a O9.5V star. However, 
this peculiar wind could explain the observed behavior in the X-ray ranges if wind-fed accretion takes 
place in an moderately eccentric (e$\sim$0.15) orbit.

\begin{figure}[h]
\centering
\includegraphics[width=15cm]{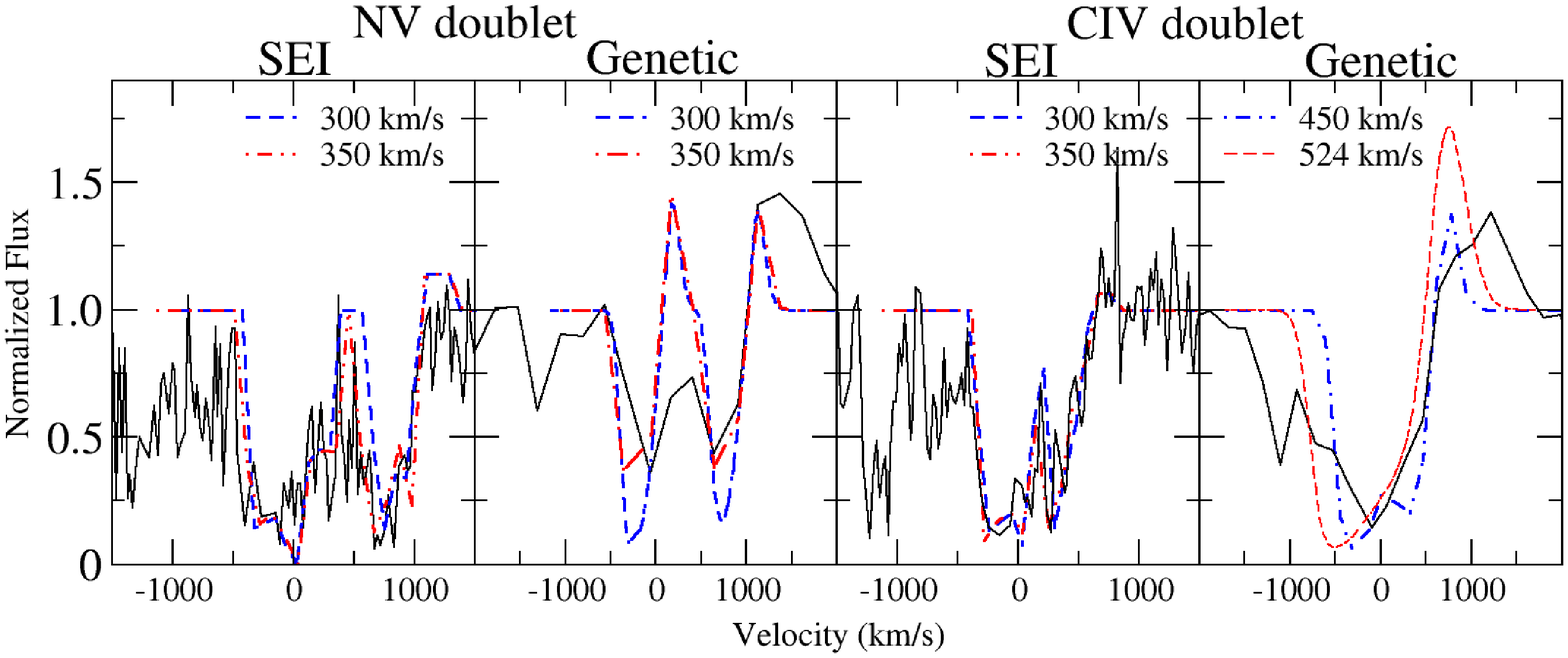}
\caption{ Fit to the observed NV and CIV doublets profiles by using a genetic algorithm and a SEI code as 
explained in the text. All fits yield terminal wind velocities in the range 300-500 kms$^{-1}$. The two 
fits shown in each plot represent the lower and upper limits of the range of possible solutions. See 
Rib\'o et al. 2006.\label{fig_3}}
\end{figure}

Figure \ref{fig_3} shows the result of fitting the IUE high resolution spectrum by using a SEI code 
(Sobolev with Exact Integration, Lamers, Cerruti-Sola, \& Perionotto  1987) and a genetic algorithm developed by Georgiev 
\& Hernández (2005). Terminal wind velocities in the range 300-500 kms$^{-1}$ are found, as explained 
in more detail in Rib\'o et al. (2006) and Blay et al. (2009).

\section{Optical data: a very intriguing star.}
\begin{figure}[h]
\begin{minipage}{10cm}
\centering
\includegraphics[width=9cm]{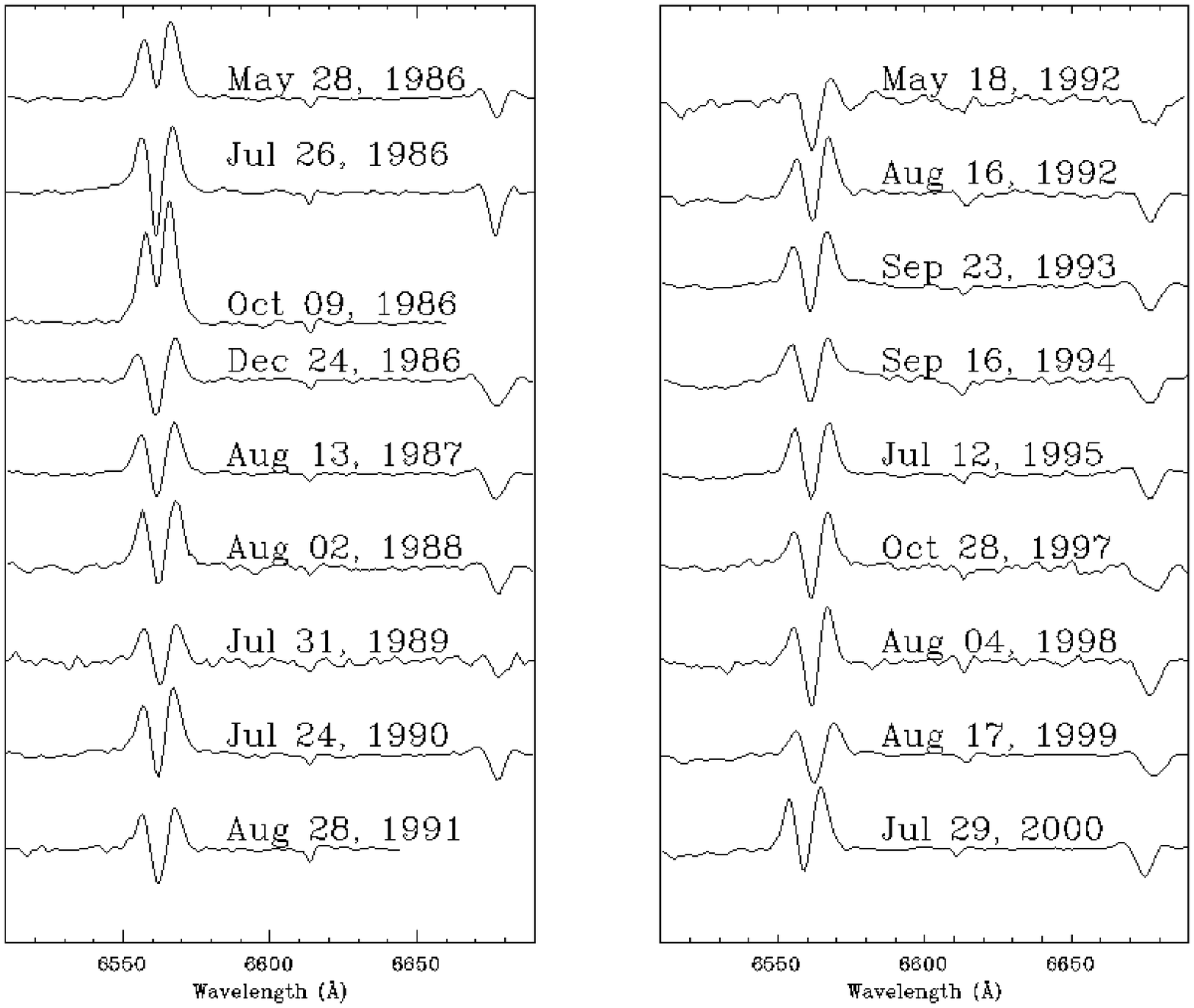}
\end{minipage}
\hfill
\begin{minipage}{6cm}
\centering
\includegraphics[width=5cm]{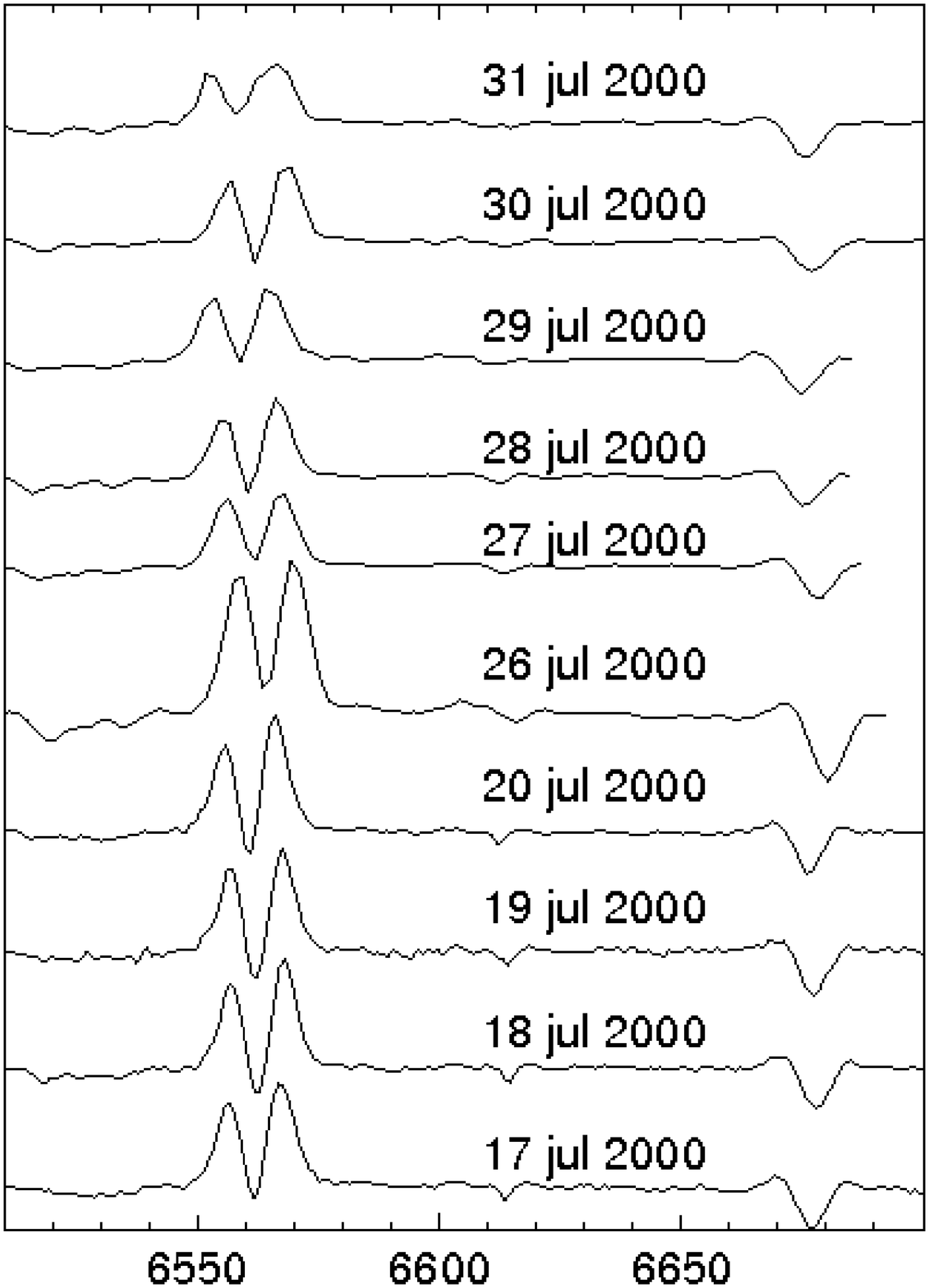}
\end{minipage}
\caption{H$\alpha$ line variability on long time scales (left panels) and shorter time scales (right panel). See Blay et al. (2006).\label{fig_4}}
\end{figure}

\begin{figure}[h]
\centering
\includegraphics[width=12cm]{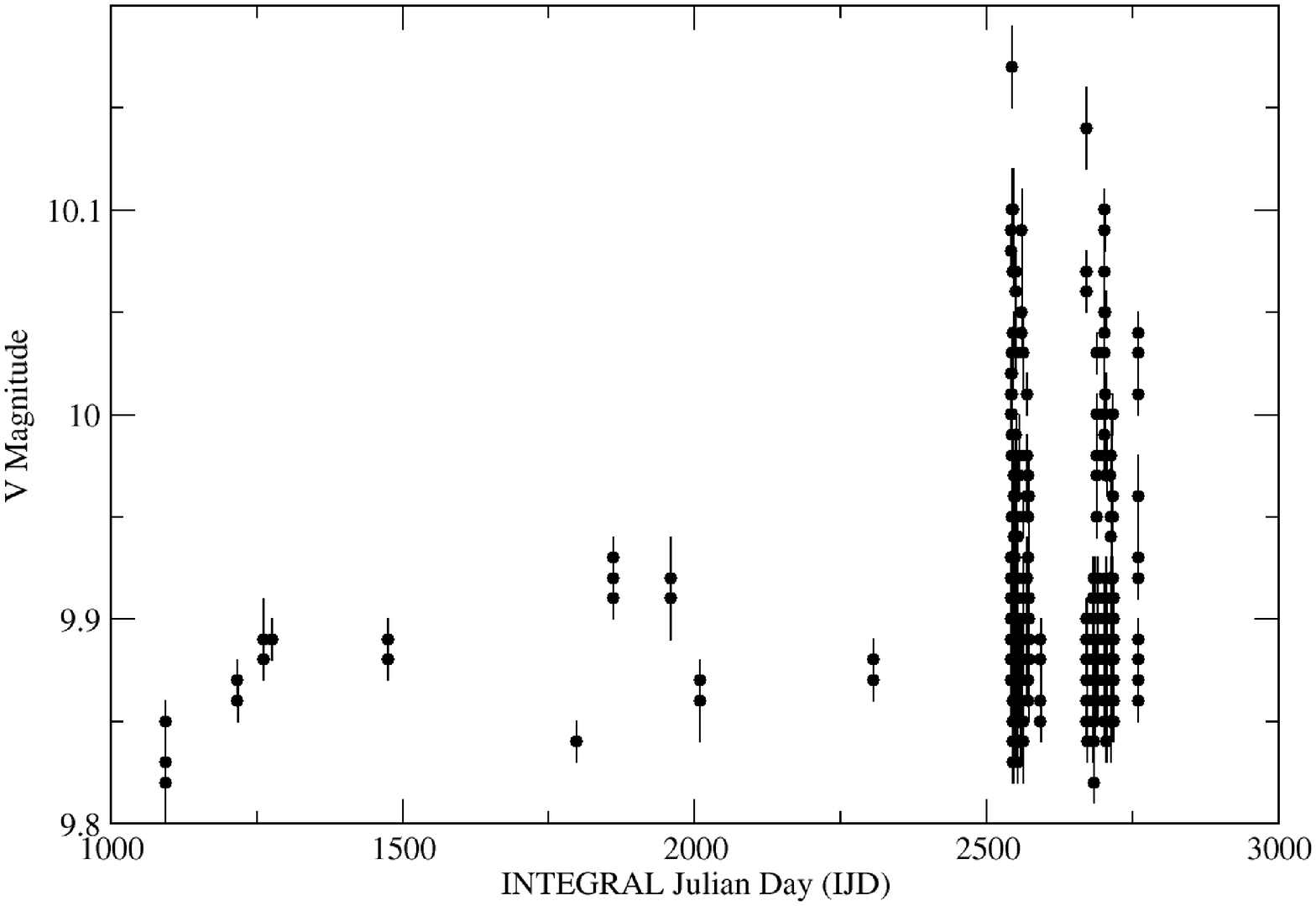}
\caption{V magnitude observations of the OMC (Optical Monitoring Camera) on board the high energy satellite INTEGRAL. The OMC works in the V filter range. Large amplitude variability is seen at some epochs, however, in the long term, the V magnitude average remains fairly constant.\label{fig_5}}
\end{figure}

Negueruela \& Reig (2001) and Blay et al. (2006) describe in very fine detail the optical peculiarities of BD +53$^\circ$2790. The adopted spectral 
type and luminosity for this massive star are O9.5Vp. There are still some peculiarities to be explained in the optical 
spectra of BD +53$^\circ$2790: a) the presence of metallic lines which is reminiscent of  those of later types (see Negueruela \& Reig, 2001); 
b) when compared to standard stars of the same or very similar spectral type, BD +53$^\circ$2790 seems to be overabundant in He (Blay et al. 2006 
suggest the possibility to link BD +53$^\circ$2790 to the group of He-rich stars). When comparing the He profiles to those of standard stars 
which are  broadened rotationally, a rotational velocity of ~315 kms$^{-1}$ is found for this object. Figure \ref{fig_4} depicts the variability 
of the H$\alpha$ line. Although the double peaked profile is always present, with the red peak always higher than the blue one, the variability is 
evident on short and long time scales. V magnitude variability is shown in Figure \ref{fig_5}. The average V magnitude remains fairly constant over 
time, but on shorter time scales it is clearly variable, with no coherent variability detected so far.

\section{IR data: The cirsumstellar environment}
\begin{figure}[h]
\centering
\includegraphics[width=12cm]{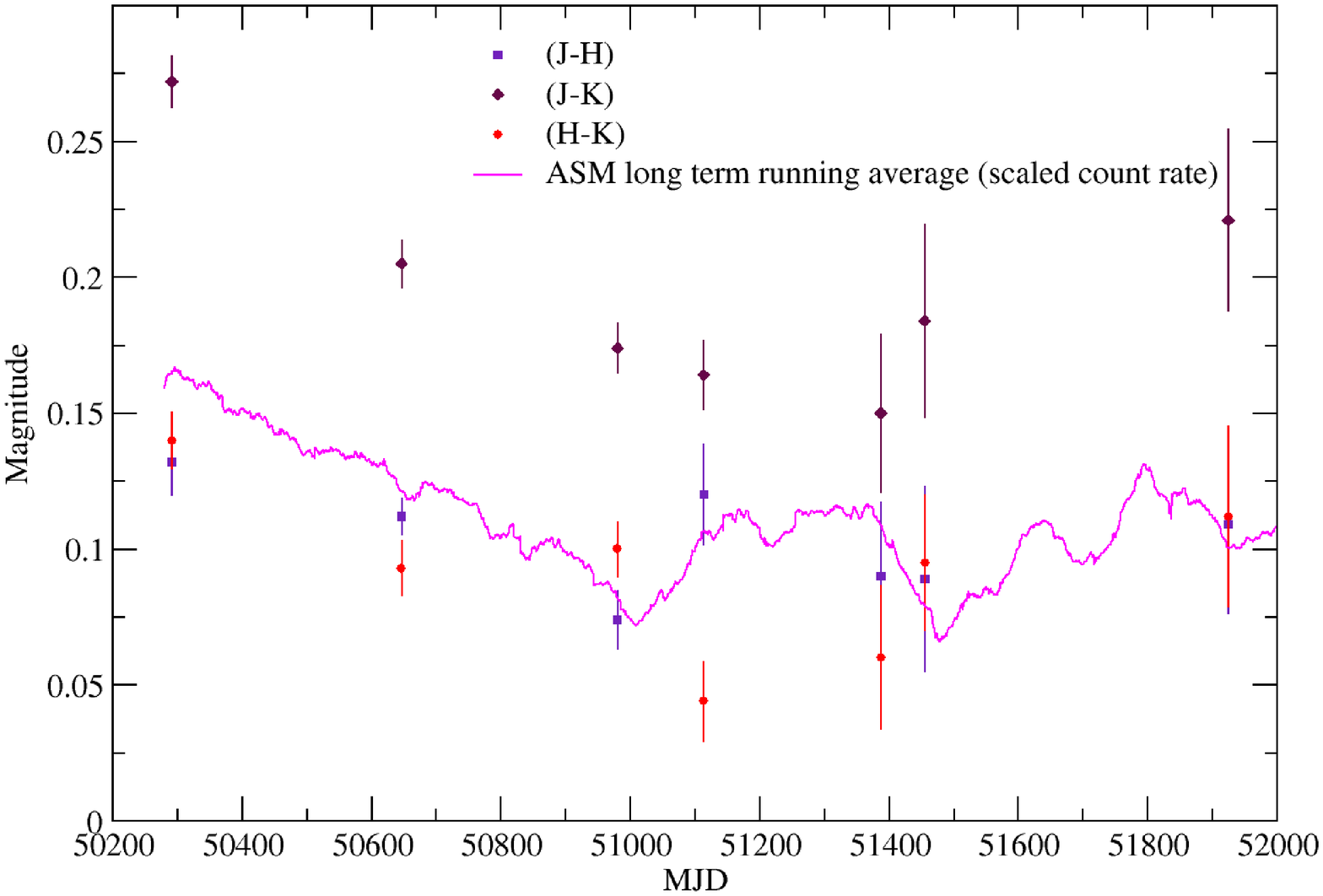}
\caption{Comparison of the long term IR colours (J-H), (J-K) and (H-K) with the long-term trend of the RXTE/SAM count rate as a running average over 300 days. RXTE/ASM count rate has been scaled for plotting purposes.\label{fig_6}}
\end{figure}

Infra-Red magnitudes and colors obtained during a long-term monitoring campaign (1987-2001) are described and analyzed
in Blay et al. (2006). No coherent variability is shown by the IR data. However a long-term trend seems to be 
present in the J and H magnitudes, but not shared by the K magnitude. On the other hand, as shown in Figure \ref{fig_6}, 
the (J-H), (J-K) and (H-K) colors show a long-term trend  which seems to be  very well correlated with 
RXTE/ASM X-ray data (in the 2-10keV energy range) in the case of the two former colors,  but not so well 
correlated when the latter colour is considered. This is an indication that X-ray emission could be 
proportional to the  IR excess, i.e., to the amount of available circumstellar matter. Spectroscopic classification
in the IR band indicates a more evolved nature of BD+53$^\circ$2790 (see Negueruela \& Reig, 2001).

\section{Radio data: as expected for a neutron star.}
Radio emission from 4U 2206+54 was analyzed in Blay et al. (2005). An image was obtained with the VLA at 
8.4 GHz after concatenation of observations from 2 runs during May 2003, with no radio emission detected at a significant level. 
According to Gallo, Fender, \& Pooley (2003) the presence 
of a Black  Hole would imply the emission of radio at the range 4-15 GHz. As stated by Blay et al. (2005), the expected
 radio emission from 4U 2206+54 (if hosting a black hole) would be 60 times greater than the 3$\sigma$ upper limit given by the VLA observation.


\section{Conclusions}
Given the peculiarity of both of its components, 4U 2206+54 shows up as a unique type of X-Ray binary system. There are only 
two known High Mass X-Ray  Binary systems showing wind-fed accretion from a main sequence star. One of them is 4U 2206+54, 
the other system is the microquasar LS 5039, which hosts a black hole and a O6.5V star. These systems are of great importance 
as they are presumably the progenitors of Supergiant HMXRBs. 4U 2206+54 also shows similarities with the supergiant systems 2S 0115+54, 
which also hosts a slowly rotating Neutron Star (2.78 h), and IGR J16358-4726, with a 1.6 h X-ray pulsar. BD+53$^\circ$2790 is a very 
active peculiar star, possibly belonging to the class of He rich stars. It very likely hosts a  magnetically driven equatorial disk-like 
envelope, but its existence is still to be confirmed by the detection of intense magnetic field, by magnitude variations modulated by 
its rotational motion, or both.

%
%
\section*{Acknowledgements}
This research is supported by the Spanish MCINN, poject 20071152/ASIM-GRI.
This work is a compilation of the results from the collaboration with I. Negueruela, P. Reig, M. Rib\'o 
and a large number of observers. 
%
%
\footnotesize
\beginrefer

\refer Blay, P., Rib\'o M., Negueruela, I., Torrej\'on, J. M., Reig, P., Camero, A., Mirabel, I. F., \& Reglero, V., 2005, A\&A, 438, 963

\refer Blay, P., Negueruela, I., Reig, P., Coe, M. J.,; Corbet, R. H. D., Fabregat, J., \& Tarasov, A. E., 2006, A\&A, 446, 1095

\refer Blay, P., Rib\'o, M.,\&  Negueruela I., 2009, Ap\&SS, 320, 145

\refer Corbet, R.H.D., Markwardt, C.B., \& Tueller, J. 2007, ApJ, 655, 458 

\refer  Finger, Mark H., Ikhsanov, Nazar R., Wilson-Hodge, C. A.,  \& Patel, S. K., 2010, ApJ, 709, 1249

\refer Gallo, E., Fender, R.E., \& Pooley, G.G., 2003, MNRAS, 344, 60

\refer Georgiev, L., Hernández, X., 2005, RMAA, 41, 121

\refer Lamers, H.J.G.L.M., Cerruti-Sola, M., \& Perionotto, M., 1987, ApJ, 314, 726

\refer Negueruela, I., \& Reig, P., 2001, A\&A, 371, 1056

\refer Reig, P., Torrej\'on, J. M., Negueruela, I., Blay, P., Rib\'o, M., \& Wilms, J., 2009, A\&A, 494, 1073

\refer Rib\'o, M., Negueruela, I., Blay, P., Torrej\'on, J. M., \& Reig, P., 2006, A\&A, 687, 698

\endrefer           
\end{document}